\pdfoutput=1

\documentclass[11pt,twoside,a4paper,cmspaper,final,collab]{cms-tdr}
\def\svnVersion{470061P}\def\cmsCernNoTag{CERN-EP-2018-136}\def\cmsCernDate{\today}\def\cmsMessage{Published in the Journal of High Energy Physics as \href{http://dx.doi.org/10.1007/JHEP07(2018)115}{\doi{10.1007/JHEP07(2018)115.}}}

\begin{document}\cmsNoteHeader{EXO-17-029}

\hyphenation{had-ron-i-za-tion}
\hyphenation{cal-or-i-me-ter}
\hyphenation{de-vices}
\RCS$Revision: 469408 $
\RCS$HeadURL: svn+ssh://svn.cern.ch/reps/tdr2/papers/EXO-17-029/trunk/EXO-17-029.tex $
\RCS$Id: EXO-17-029.tex 469408 2018-07-23 13:48:21Z ytakahas $

\newcolumntype{P}[1]{>{\centering\arraybackslash}p{#1}}
\newcommand{\MT}{\ensuremath{m_\mathrm{T}}\xspace}
\newcommand{\emu}{\Pe\Pgm}
\newcommand{\mutau}{\Pgm\tauh}
\newcommand{\etau}{\Pe\tauh}
\newcommand{\tautau}{\tauh\tauh}
\newcommand{\ltau}{\ell\tauh}
\newcommand{\mvis}{m_\text{vis}}
\newcommand{\st}{S_\text{T}}
\newcommand{\mLQ}{m_\text{LQ}}
\newcommand{\zjet}{\PZ\text{$+$jets}}
\newcommand{\wjet}{\PW\text{$+$jets}}

\newlength\cmsTabSkip\setlength{\cmsTabSkip}{1ex}

\cmsNoteHeader{EXO-17-029}

\title{Search for a singly produced third-generation scalar leptoquark decaying to a $\tau$ lepton and a bottom quark in proton-proton collisions at $\sqrt{s} = 13$\TeV}

\date{\today}

\abstract{
        A search is presented for a singly produced third-generation scalar leptoquark decaying to a $\tau$ lepton and a bottom quark. Associated production of a leptoquark and a $\tau$ lepton is considered, leading to a final state with a bottom quark and two $\tau$ leptons. The search uses proton-proton collision data at a center-of-mass energy of 13\TeV recorded with the CMS detector, corresponding to an integrated luminosity of 35.9\fbinv. Upper limits are set at 95\% confidence level
on the production cross section of the third-generation scalar leptoquarks as a function of their mass.
From a comparison of the results with the theoretical predictions, a third-generation scalar leptoquark decaying to a $\tau$ lepton and a bottom quark, assuming unit Yukawa coupling ($\lambda$), is excluded for masses below 740\GeV.
Limits are also set on $\lambda$ of the hypothesized leptoquark as a function of its mass.
Above $\lambda = 1.4$, this result provides the best upper limit on the mass of a third-generation scalar leptoquark decaying to a $\tau$ lepton and a bottom quark.
}

\hypersetup{
pdfauthor={CMS Collaboration},
pdftitle={Search for a singly produced third-generation scalar leptoquark decaying to a tau lepton and a bottom quark in proton-proton collisions at sqrt(s) = 13 TeV},
pdfsubject={CMS},
pdfkeywords={CMS, physics, leptoquark, third-generation}}

\maketitle

\section{Introduction}
\label{sec:intro}
Leptoquarks (LQs) are hypothetical color-triplet bosons, which carry both baryon and lepton quantum numbers and have fractional electric charge. They are predicted by many extensions of the standard model (SM) of particle physics, such as theories invoking grand unification\,\cite{Georgi:1974sy,Pati:1973uk,Pati:1974yy,Murayama:1991ah,Fritzsch:1974nn,Senjanovic:1982ex,Frampton:1989fu,Frampton:1990hz}, technicolor\,\cite{Dimopoulos:1979es,Dimopoulos:1979sp,Farhi:1980xs}, or compositeness\,\cite{Schrempp:1984nj}.
To satisfy experimental constraints on flavor changing neutral currents and other rare processes\,\cite{Buchmuller:1986iq,Shanker:1982nd}, it is generally assumed that there would be
three types of LQs, each type coupled to leptons and quarks of its same generation.

Third-generation scalar LQs have recently received considerable theoretical interest,
as their existence can explain the anomaly in the $\overline{\PB} \to \PD \tau \overline{\nu}$ and
$\overline{\PB} \to \PD^{*} \tau \overline{\nu}$ decay rates reported by the
BaBar\,\cite{Lees:2012xj, Lees:2013uzd}, Belle\,\cite{Matyja:2007kt, Bozek:2010xy, Huschle:2015rga, Sato:2016svk, Hirose:2016wfn, Hirose:2017dxl},
and LHCb\,\cite{Aaij:2015yra} Collaborations. These decay rates collectively deviate from the SM predictions by about
four standard deviations\,\cite{Dumont:2016xpj}, and
large couplings to third-generation quarks and leptons could explain this anomaly\,\cite{Tanaka:2012nw, Sakaki:2013bfa, Dorsner:2013tla, Gripaios:2014tna}. The LQ could also provide consistent explanations for other anomalies in $\PB$ physics reported by LHCb\,\cite{Aaij:2014pli,Aaij:2016flj,Aaij:2015esa,Aaij:2015oid,Aaij:2014ora,Aaij:2017vbb} and Belle\,\cite{Wehle:2016yoi}.

The production cross sections and decay widths of LQs in proton-proton ($\Pp\Pp$) collisions are determined by the LQ's mass, $\mLQ$; its branching fraction $\beta$ to a charged lepton and a quark; and the Yukawa coupling $\lambda$ of the LQ-lepton-quark vertex.
Leptoquarks can be produced in pairs via gluon fusion or quark-antiquark annihilation, and singly via quark-gluon fusion. Pair production of LQs does not depend on $\lambda$, while single production does, and thus the sensitivity of searches for singly-produced LQs depends on $\lambda$. At lower masses, the cross section for pair production is greater than that for single production. However, the single-LQ production cross section decreases more slowly with increasing $\mLQ$,
eventually exceeding that for pair production.
If the third-generation LQ is responsible for the observed $\PB$ physics anomalies, then a large value of $\lambda$ is favored ($\lambda \sim \mLQ$ measured in \TeV), and the cross section for single production exceeds that for pair production for $\mLQ$ greater than 1.0–-1.5\TeV\,\cite{Buttazzo:2017ixm}.

The most stringent limits on the production of a third-generation LQ decaying to a $\tau$ lepton and a bottom quark comes from a search by the CMS Collaboration, in which a scalar LQ with mass below 850\GeV was excluded in a search for LQ pair production in the $\ltau\PQb\PQb$ final state\,\cite{Sirunyan:2017yrk}. Here, $\ell$ refers to a lepton ($\Pe$ or $\Pgm$) from $\tau$ lepton decay (35\% of the $\tau$ decays\,\cite{Patrignani:2016xqp}), and $\tauh$ denotes a hadronically decaying $\tau$ lepton (65\% of the $\tau$ decays\,\cite{Patrignani:2016xqp}).
Another type of third-generation scalar LQ decaying to a $\tau$ lepton and a top quark is excluded for masses up to 900\GeV\,\cite{Sirunyan:2018nkj}.

\begin{figure}[h!t]
\centering
\includegraphics[width=0.40\textwidth]{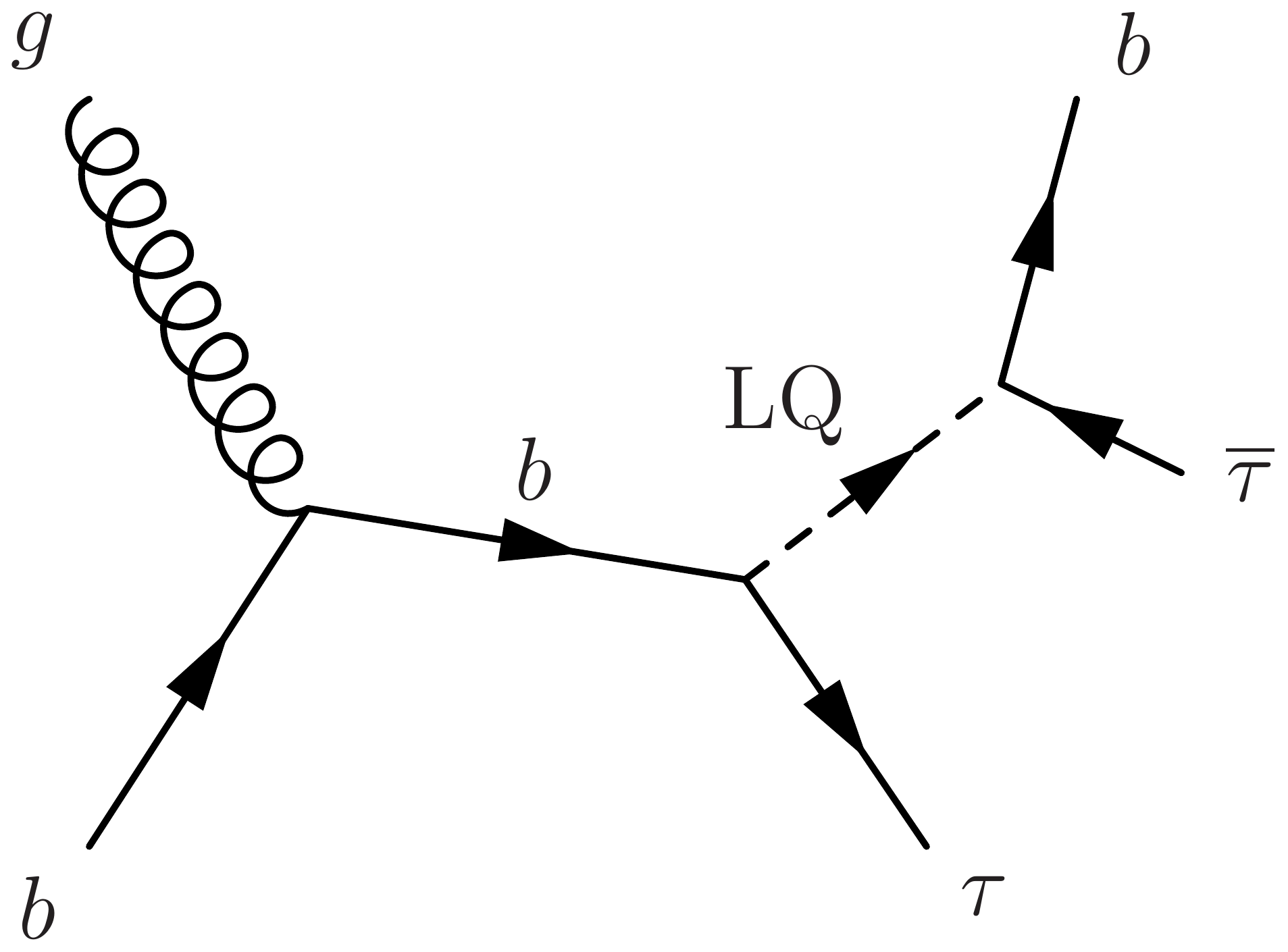}
\hspace{0.5cm}
\includegraphics[width=0.35\textwidth]{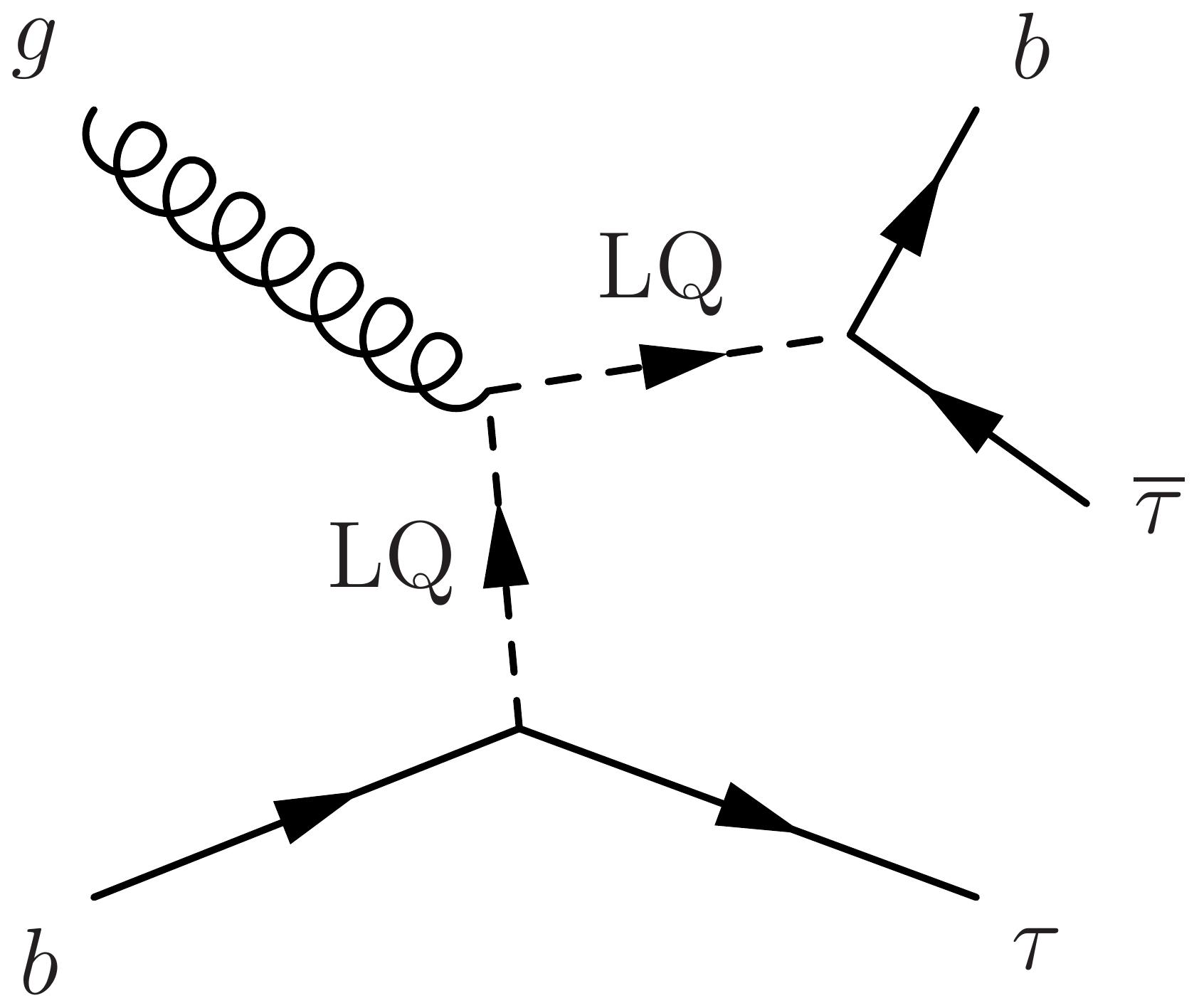}
\caption{Leading order Feynman diagrams for the single production of third-generation LQs subsequently decaying to a $\tau$ lepton and a bottom quark, for the $s$-channel (left) and $t$-channel (right) processes.}
\label{diagram}
\end{figure}

This paper presents the first search that targets singly produced third-generation scalar LQs, each decaying to a $\tau$ lepton and a bottom quark.
Feynman diagrams of the signal processes at leading order (LO) are shown in Fig.\,\ref{diagram}.
The final states $\ltau$b and $\tautau\PQb$ are considered. The search is based on a data sample of $\Pp\Pp$ collisions at a center-of-mass energy of 13\TeV recorded
by the CMS experiment at the CERN LHC in 2016, corresponding to an integrated luminosity of 35.9\fbinv.

\section{The CMS detector} \label{sec:dectector}

The central feature of the CMS apparatus is a superconducting solenoid of 6\unit{m} internal diameter, providing a magnetic field of 3.8\unit{T}. Within the solenoid volume, there are a silicon pixel and strip tracker, a lead tungstate crystal electromagnetic calorimeter (ECAL), and a brass and scintillator hadron calorimeter, each composed of a barrel and two endcap sections. Forward calorimeters extend the pseudorapidity coverage provided by the barrel and endcap detectors. Muons are measured in gas-ionization detectors embedded in the steel flux-return yoke outside the solenoid.

Events of interest are selected using a two-tiered trigger system\,\cite{Khachatryan:2016bia}. The first level, composed of custom hardware processors, uses information from the calorimeters and muon detectors to select events at a rate of around 100\unit{kHz} within a time interval of less than 4\mus. The second level, known as the high-level trigger, consists of a farm of processors running a version of the full event reconstruction software optimized for fast processing, and reduces the event rate to about 1\unit{kHz} before data storage.

A more detailed description of the CMS detector, together with a definition of the coordinate system used and the relevant kinematic variables, can be found in Ref.\,\cite{Chatrchyan:2008zzk}.

\section{Simulated samples}

Samples of simulated events are used to devise selection criteria and to estimate and validate background predictions.
The LQ signals are generated at LO precision using version 2.6.0 of $\MGvATNLO$\,\cite{aMCatNLO} for $\mLQ =$ 200--1000 (in steps of 100\GeV), 1200, and 1500\GeV. The particular LQ model used is $\tilde{R}_2$, as discussed in Ref.\,\cite{Dorsner:2016wpm}. The branching fraction is assumed to be $\beta=1$, \ie the LQ\ always decays to a $\tau$ lepton and a bottom quark. The Yukawa coupling of the LQ to a $\tau$ lepton and a bottom quark is set to be $\lambda=1$. The width $\Gamma$ is calculated to be $\Gamma = \mLQ \lambda^2 / (16\pi)$\,\cite{Plehn:1997az}, which is narrower than the experimental resolution over the considered search range. The signal samples are normalized to the cross section calculated at LO precision, multiplied by a $K$ factor to account for higher order contributions\,\cite{Dorsner:2018ynv}. The $K$ factors are almost constant as a function of $\mLQ$ and are approximately 1.4 for the bottom-quark-initiated diagrams considered in this analysis.

The main sources of background are the pair production of top quarks ($\ttbar$), $\PW$ and $\PZ$ boson production in association with jets, denoted as ``$\wjet$'' and ``$\zjet$'', diboson ($\PW\PW, \PW\PZ, \PZ\PZ$), single production of top quarks, and quantum chromodynamics (QCD) production of multijet events. The $\wjet$ and $\zjet$ processes are simulated using the $\MGvATNLO$ generator (v5\_2.2.2 and v5\_2.3.3) at LO precision with the MLM jet matching and merging scheme\,\cite{Alwall:2007fs}.
The same generator is also used for diboson production simulated at next-to-leading order (NLO) with the FxFx jet matching and merging scheme\,\cite{Campbell:2011bn,Frederix:2012ps}, {\tolerance=800 whereas $\POWHEG$\,\cite{Nason:2004rx,Frixione:2007vw, Alioli:2010xd, Alioli:2010xa} 2.0 and 1.0 are used for $\ttbar$ and single top quark production at NLO precision, respectively\,\cite{Frixione:2007nw,Alioli:2009je,Re:2010bp}. The $\ttbar$ process is normalized using cross sections calculated at next-to-next-to-leading order in perturbative QCD\,\cite{Czakon:2011xx}.\par}

The generators are interfaced with \PYTHIA 8.212\,\cite{Sjostrand:2014zea} to model the parton showering and fragmentation, as well as the decay of the $\Pgt$ leptons. The \PYTHIA parameters affecting the description of the underlying event are set to the {CUETP8M1} tune\,\cite{Khachatryan:2015pea}. The NNPDF3.0 parton distribution functions\,\cite{Ball:2011uy,Ball:2014uwa} with the QCD order matching that of the matrix element calculations are used with all generators.

Simulated events are processed with a model of the CMS detector based on \GEANTfour\,\cite{Agostinelli:2002hh} and are reconstructed with the same algorithms used for data. The effect of pileup, additional $\Pp\Pp$ collisions within the same or adjacent bunch crossings, is taken into account by adding minimum bias events, generated with \PYTHIA, to the hard scattering event. The additional events are weighted such that the frequency distribution matches that in data, with an average of approximately 23 interactions per bunch crossing\,\cite{CMS-PAS-LUM-17-001}.

\section{Event reconstruction}
\label{sec:reconstruction}

The reconstruction of observed and simulated events uses a particle-flow (PF) algorithm\,\cite{Sirunyan:2017ulk},
which combines the information from the CMS subdetectors to identify
and reconstruct the particles emerging from $\Pp\Pp$ collisions:
charged and neutral hadrons, photons, muons, and electrons.
Combinations of these PF objects are used to reconstruct
higher-level objects such as jets, $\tauh$ candidates, or
missing transverse momentum ($\ptvecmiss$), taken as the negative vector sum of the transverse momenta (\pt) of those jets.

The reconstructed vertex with the largest value of summed physics-object $\pt^2$ is taken to be the primary $\Pp\Pp$ collision vertex. In this case, the physics objects are the objects constructed by a jet finding algorithm\,\cite{Cacciari:2008gp,Cacciari:fastjet1} applied to all charged tracks associated with the vertex, including tracks from lepton candidates, and the corresponding associated $\ptvecmiss$.

Electrons are identified with a multivariate (MVA)\,\cite{Hocker:2007ht} discriminant
combining several quantities describing the track quality,
the shape of the energy deposits in the ECAL,
and the compatibility of the measurements from the tracker and the
ECAL\,\cite{Khachatryan:2015hwa}. The electrons must pass a cut-based discriminant to reject electrons coming from photon conversions.
Muons are identified with requirements on the quality of
the track reconstruction and on the number of measurements in the
tracker and the muon systems\,\cite{Chatrchyan:2012xi}.
Electron and muon candidates are required to have $\pt > 50$\GeV.
To reject leptons that do not come from the primary vertex and particles misidentified as leptons, a relative lepton isolation $I^{\ell}$ ($\ell = \Pe, \mu$) is defined as follows:
\begin{equation}
I^{\ell} \equiv \frac{\sum_\text{ch}  \PT + \max\left( 0, \, \sum_\text{neut}  \PT
                                         - \frac{1}{2} \sum_\text{ch, PU} \PT  \right )}{\PT^{\ell}}. \notag
\label{eq:reconstruction_isolation}
\end{equation}
In this expression, $\sum_\text{ch}  \PT$ is the scalar sum of the
$\pt$ of the charged hadrons, electrons, and muons originating from
the primary vertex and located in a cone of size
\mbox{$\Delta R = \sqrt{\smash[b]{(\Delta \eta)^2 + (\Delta \phi)^2}}$} \mbox{$= 0.03$} ($0.04$) centered on the electron (muon) direction, where $\eta$ is pseudorapidity and $\phi$ is azimuthal angle in radians. The sum
$\sum_\text{neut}  \PT$  represents
the same quantity for neutral hadrons and photons.
The contribution of pileup photons
and neutral hadrons is estimated from the scalar sum of the $\pt$ of charged hadrons originating from pileup vertices, $\sum_\text{ch, PU} \PT$. This sum is multiplied by a factor of
$1/2$, which corresponds approximately to the ratio of neutral- to charged-hadron production in the hadronization process
of inelastic $\Pp\Pp$ collisions, as estimated from simulation.
In this analysis, $I^{\Pe} < 0.10$ ($I^{\mu} < 0.15$) is used as an isolation requirement for the electron (muon).
With these cut-off values, the combined efficiency of identification and isolation is around 80 (95)\% for the electron (muon).
Small differences, up to the 5\% level, between data and simulation are corrected for by applying scale factors to simulated events.

Jets are reconstructed from PF candidates using an anti-\kt clustering algorithm with a distance parameter of 0.4, implemented in the \FASTJET library\,\cite{Cacciari:fastjet1, Cacciari:fastjet2}.
Charged PF candidates
not associated with the primary vertex of the interaction
are not considered when reconstructing jets.
An offset correction is applied to jet energies to take into account the contribution from additional $\Pp\Pp$ collisions within the same or nearby bunch crossings.
Jet energy corrections are derived from simulation to bring the average measured response of jets to that of particle-level jets\,\cite{CMS-JME-10-011}.
Further identification requirements are applied to distinguish genuine jets from those coming from pileup\,\cite{CMS-PAS-JME-16-003, CMS-PAS-JME-13-005}. In this analysis, jets are required to
have $\pt$ greater than 30\GeV and $\abs{\eta}$ less than 4.7, and must be separated from the selected leptons by a $\Delta R$ of at least 0.5.
Jets originating from the hadronization of bottom quarks are identified ($\PQb$ tagged)
using the combined secondary vertex algorithm\,\cite{Sirunyan:2017ezt},
which exploits observables related to the long lifetime and large mass of $\PB$ hadrons.
The chosen $\PQb$ tagging working point corresponds to an identification efficiency of approximately 60\% with
a misidentification rate of approximately 1\%, for jets originating from light (up, down, charm, strange) quarks and gluons.

The $\tauh$ candidates are reconstructed with the hadron-plus-strips
algorithm\,\cite{Khachatryan:2015dfa, CMS-PAS-TAU-16-002}, which is
seeded with anti-\kt jets.
This algorithm reconstructs $\tauh$ candidates in the one-prong,
one-prong + $\pi^0$(s), and three-prong decay modes,
based on the
number of tracks and on the number of strips of ECAL crystals with energy deposits.
An MVA-based discriminator, including isolation
as well as lifetime information, is used to reduce the incidence of jets
being misidentified as $\tauh$ candidates.
The typical working point used in this analysis has an efficiency ${\approx}60$\% for a genuine $\tauh$, with a misidentification rate for quark and gluon jets of ${\approx}0.1$\%.

Electrons and muons misidentified as $\tauh$ candidates are suppressed using criteria
based on the consistency between the measurements in the tracker, the calorimeters, and the muon detectors.
The criteria are optimized separately for each final state studied.

All particles reconstructed in the event are used in the determination of \ptvecmiss\,\cite{CMS-PAS-JME-12-002}.
The calculation takes into account jet energy corrections.
Corrections are applied to correct for the mismodeling of $\ptvecmiss$ in the simulated samples
of the $\zjet$ and $\wjet$ processes.
The corrections are performed on the variable defined as the vectorial difference
between the measured $\ptvecmiss$ and the total $\pt$ of neutrinos
originating from the decay of the $\PW$ or $\PZ$ boson.

\section{Event selection}
\label{sec:selection}

The search for scalar LQs is performed in three channels, each containing a $\PQb$-tagged jet. Channels containing in addition an electron or a muon, together with a $\tauh$ candidate are labeled $\etau$ and $\mutau$, and collectively referred to as $\ltau$. The third channel, which has two $\tauh$ candidates in addition to the $\PQb$-tagged jet, is labeled $\tautau$.
The selection criteria have been optimized based on the expected sensitivity to a single-LQ signal.

Firstly, events are required to be compatible with $\tau\tau$ production.
In the $\etau$ ($\mutau$) channel, events are selected using a trigger that requires an isolated electron (muon) with $\pt > 25$ ($24$)\GeV. Offline, the selected electron (muon) is required to have $\pt > 50\GeV$ and $\abs{\eta} < 2.1$ (2.4). The $\tauh$ candidate is required to have $\pt > 50\GeV$ and $\abs{\eta} < 2.3$. In the $\tautau$ channel, events are selected online by requiring two isolated $\tauh$ candidates with $\pt > 35\GeV$. Offline, both $\tauh$ candidates are required to have $\pt > 50\GeV$ and $\abs{\eta} < 2.1$. The selected $\ell$ and $\tauh$ candidate, or two $\tauh$ candidates, must meet isolation requirements as detailed in Section\,\ref{sec:reconstruction}, have opposite-sign (OS) electric charges and be separated by $\Delta R>0.5$. They must also meet the requirement that the distance of closest approach to the primary vertex satisfies $\abs{d_z}<0.2\unit{cm}$ along the beam direction, and $\abs{d_{xy}}<0.045\unit{cm}$ in the transverse plane.
Events with additional isolated muons or electrons ($\pt > 10\GeV$ and $\abs{\eta} < 2.4$ or 2.5) that pass a looser identification requirement are discarded to reduce $\zjet$ and diboson backgrounds and to avoid correlations between channels.

Further event selection is applied to increase the signal purity.
Since signal events contain at least one energetic bottom quark jet coming from the LQ decay,
at least one $\PQb$-tagged jet with $\pt > 50\GeV$ and $\abs{\eta} < 2.4$ is required.
To reduce the $\zjet$ background, the invariant mass, $\mvis$, of the $\ell$ and $\tauh$ candidate (two $\tauh$ candidates), is required to be greater than 85 (95)\GeV in the $\ltau$ ($\tautau$) channels.

\begin{figure}[h!t]
\centering
\includegraphics[width=0.6\textwidth]{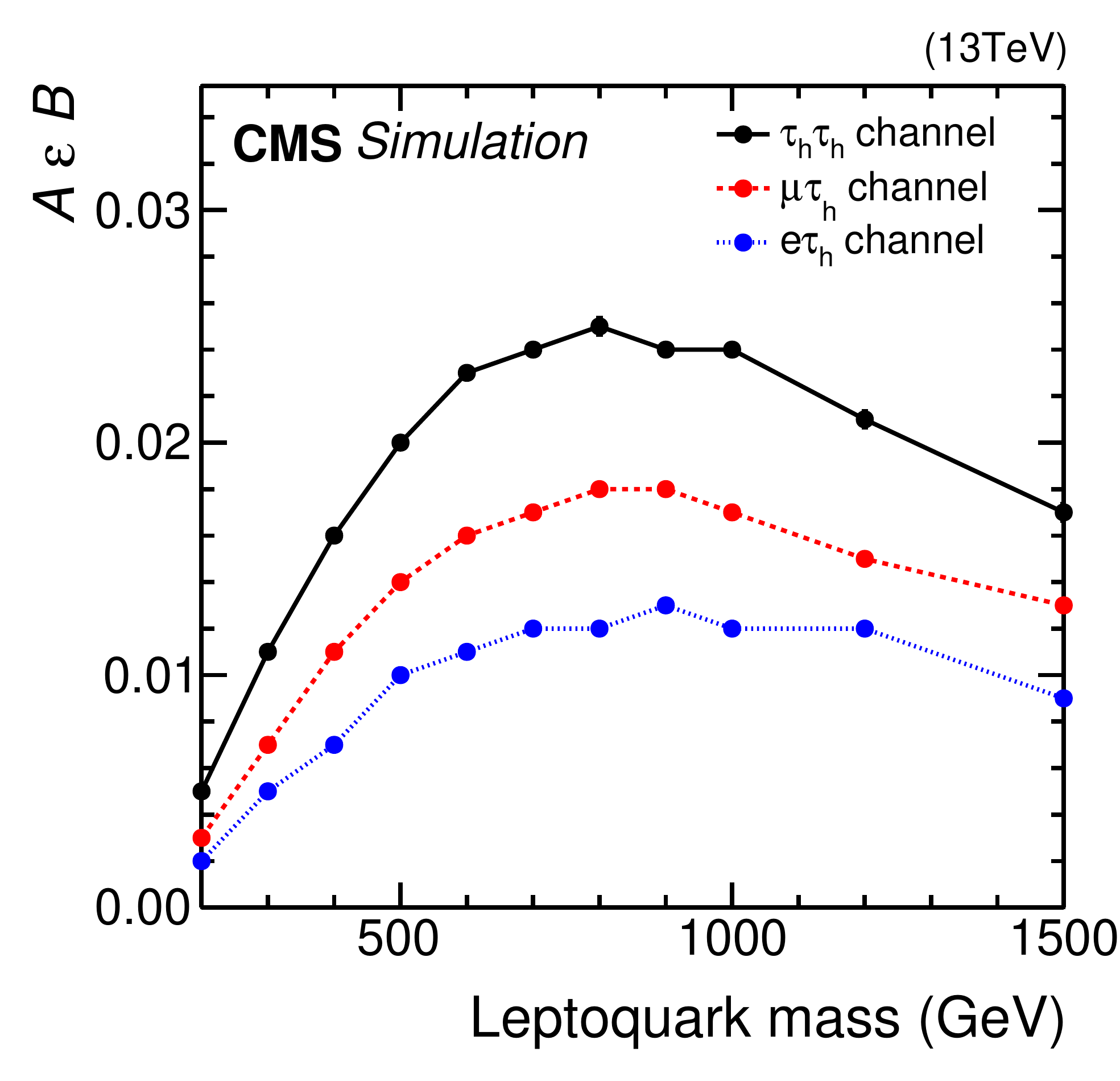}
\caption{The product of acceptance, efficiency, and branching fraction as a function of $\mLQ$ for the single production of LQs in each of the three channels considered: $\tautau$ (black solid line), $\mutau$ (red dashed line), and $\etau$ (blue dotted line).}
\label{fig:acc}
\end{figure}

The product of acceptance (${\cal A}$), efficiency ($\varepsilon$), and the branching fraction (${\cal B}$) of the $\tau\tau$ to a specific final state ranges from 0.2 to 1.3\% in the \etau channel for $\mLQ$ between 200 and 1500\GeV. The $\varepsilon$ increases with increasing $\mLQ$ due to the harder $\pt$ spectra of the final state particles.
Beyond 1000\GeV, however, $\varepsilon$ starts to decrease, mainly because of the lower $\PQb$ tagging efficiency.
Similarly, ${\cal A}\varepsilon {\cal B}$ in the $\mutau$ ($\tautau$) channels range from 0.3 to 1.8\% (0.5 to 2.5\%). Figure\,\ref{fig:acc} shows ${\cal A}\varepsilon {\cal B}$ for the signal, in each final state considered in this analysis, as a function of $\mLQ$.

After applying the event selection, an excess of events over the SM backgrounds is searched for using the distribution of the scalar $\pt$ sum of all required final-state particles, $\st$, which is defined as
$\pt(\ell) + \pt(\tauh) + \pt(\text{leading jet})$ for the $\ltau$ channels, and {\tolerance=800 $\pt(\text{leading }\tauh) + \pt(\text{subleading }\tauh) + \pt(\text{leading jet})$ for the $\tautau$ channel, where leading and subleading refer to $\pt$. Because of the $\pt$ threshold requirements, $\st$ is always greater than 150\GeV. \par}

\section{Background estimation}
\label{sec:bg}

The dominant background in all channels is $\ttbar$ production because of the presence of genuine electrons, muons, $\tau$ leptons, and bottom quark jets produced in the $\ttbar$ decays.
Additional backgrounds that satisfy the signal selection are $\wjet$, $\zjet$, diboson, and single top quark processes, as well as QCD multijet events.
In this section, background estimation methods and their validation are described separately for the $\ltau$ and $\tautau$ channels.

\subsection{The \texorpdfstring{$\ltau$}{l tau} channels}

The normalization and shape of the $\ttbar$ background are obtained from data, making use of an $\emu$ control region (CR), containing events with an electron, a muon, and at least one $\PQb$-tagged jet.
The same $\pt$ and $\abs{\eta}$ requirements as in the signal region (SR) are placed on all three objects.
The invariant mass of the selected electron and muon is required to be greater than 85\GeV.
The purity of $\ttbar$ events in this CR, estimated from simulation, is 92\%, with negligible signal contamination.
A good agreement between data and simulation is found for both the normalization and the shape,
validating the method used to estimate the $\ttbar$ background in the SR.

In order to allow for possible remaining mismodeling related to $\ttbar$ backgrounds, this CR is included in the maximum likelihood fit, as described in Section\,\ref{sec:signal_extraction}, together with relevant nuisance parameters such as the $\PQb$ tagging efficiency and \ttbar cross section uncertainties. In this way, the normalization and the shape of the $\ttbar$ backgrounds can be constrained from the data. This procedure also helps to constrain $\ttbar$ backgrounds in the $\tautau$ channel, although its contribution is less significant than in the $\ltau$ channels.

For the $\wjet$ background, the shape is taken from simulation, while the normalization is determined from data in a high ($>$80\GeV) transverse mass ($\MT$) sideband; here $\MT$ is defined as
\begin{equation}
\MT = \sqrt{{2 \pt^\ell \abs{\ptvecmiss} (1-\cos \Delta\phi)}}, \notag
\end{equation}
where $\pt^\ell$ is the lepton $\pt$ and $\Delta\phi$ is the azimuthal angle between the lepton direction and \ptvecmiss.
The normalization factor is calculated before the $\PQb$ tagging requirement is applied. A 30\% uncertainty is assigned for the $\wjet$ background estimate to account for the limited event counts in the high $\MT$ sideband,
as well as the extrapolation uncertainty to the SR.

The QCD multijet background, in which one of the jets is misidentified as the $\tauh$ candidate and another as the $\ell$, is small and is estimated
using a CR where the $\ell$ and $\tauh$ candidate have same-sign (SS) electric charges. In this CR, the QCD multijet yield is obtained by subtracting
from the data the contribution of the $\zjet$, $\ttbar$, and $\wjet$ processes.
The expected contribution of the QCD multijet background in the OS SR is then derived by rescaling the yield obtained in the SS CR by a factor of 1.06, which is measured using a pure QCD multijet
sample obtained by inverting the lepton isolation requirement.
To determine the uncertainty associated with this procedure, the measurement is repeated with several different $\st$ requirements. The maximum variation observed is 30\%, and this is taken to be the uncertainty in the QCD background estimate.

Minor backgrounds such as diboson and single top quark processes are estimated from the simulation.

\subsection{The \texorpdfstring{$\tautau$}{tau tau} channel}
In the $\tautau$ channel, the shape and normalization of all background processes with genuine hadronic $\tau$ decays are estimated using simulated samples.
The backgrounds concerned are $\PZ\to\tauh\tauh$, and contributions from the $\ttbar$, diboson and single top quark processes.

Other backgrounds arising from jets misidentified as $\tauh$ candidates, most of which are from QCD multijet backgrounds, are estimated from CRs in data using the fake-factor method\,\cite{Sirunyan:2018zut,Sirunyan:2018qio}.
An application region (AR) is defined containing the same selection criteria as in the SR, except for an inverted $\tauh$ isolation requirement for one of the two $\tauh$ candidates.
The AR is primarily populated by events with jets misidentified as $\tauh$ candidates, and has a contamination from genuine hadronic $\tau$ decays at the level of a few percent or below.

The ratio of the number of events with a misidentified $\tauh$ in the AR to the number in the SR (fake factor) is assumed to be the same as the ratio measured in samples with an SS $\tautau$ pair. The fake factor is then applied to the number of events in the AR to estimate the number of events with a misidentified $\tauh$ in the SR.
The fake factor is calculated as a function of the $\pt$ and decay mode of the $\tauh$ candidate, and it ranges from 0.1 to 0.25. In order to take combinatorial effects into account, a weight factor of 0.5 is applied. Since the presence of small backgrounds in the AR that contain a genuine $\tauh$ results in up to a 2\% underestimation of the number of events with a misidentified $\tauh$ in the SR,
a correction is applied based on the fractions of these processes in simulated events. Corresponding uncertainties are incorporated in the fit model, as described in Section\,\ref{sec:signal_extraction}.

The fake-factor method is tested in two validation regions (VRs); one is constructed by inverting the leading jet $\pt$ requirement (\ie $\pt < 50\GeV$)
and the other by using events with two $\tauh$ candidates that do not fulfill the tight isolation criteria used to define the SR.
For both VRs, all other selection criteria are kept identical to the SR, except that the $\PQb$ tagging requirement
is removed to increase the number of events. The signal contamination is negligible in both VRs. Agreement between data and simulation is found, within statistical uncertainties, demonstrating the validity of the
fake-factor method.

Finally, small contributions coming from the $\zjet$ background are validated using the same event selection as in the SR,
except that we require $\mvis < 95$\GeV.
Data and simulated events show agreement within statistical uncertainty.
A 30\% uncertainty is attributed to the $\zjet$ background yield due to the limited event count in this CR.

\section{Systematic uncertainties and signal extraction} \label{sec:signal_extraction}
A binned maximum likelihood method is used for the signal extraction\,\cite{CMS-NOTE-2011-005}.
As discussed in Section\,\ref{sec:selection}, the $\st$ distribution for $\st$ greater than 150\GeV is used as the final discriminant.
The fit is performed simultaneously in the $\etau$, $\mutau$, and $\tautau$ signal regions, as well as in the $\emu$ control region, as defined in Section\,\ref{sec:bg}.
Systematic uncertainties may affect the normalization
and the shape of the $\st$ distribution of the signal
and background processes.
These uncertainties are represented by nuisance parameters in the fit, as described below.
The relevant uncertainties are summarized in Table\,\ref{table:uncertainties}.

\subsection{Normalization uncertainties}
The uncertainty in the integrated luminosity amounts to 2.5\%\,\cite{CMS-PAS-LUM-17-001}
and affects the normalization of the signal and background processes that are based on simulation.
Uncertainties in the electron identification and trigger efficiency amount to 8 and 2\%, respectively, while those in the muon identification and trigger efficiency amount to 2\% each.
The $\tauh$ identification and trigger efficiency have been measured using
the ``tag-and-probe'' technique\,\cite{Khachatryan:2015dfa, CMS-PAS-TAU-16-002} and an uncertainty of 5\% per \tauh candidate is assigned. The acceptance uncertainty due to the $\PQb$ tagging efficiency (misidentification rate) is taken to be 3 (5)\%.
A 30\% uncertainty is attributed to the $\wjet$, $\zjet$, and QCD multijet backgrounds in the $\ltau$ channels, as discussed in Section\,\ref{sec:bg}.
The cross section uncertainties in the $\ttbar$, diboson, and single top quark processes are 5.5, 6.0, and 5.5\%, respectively.
For events where electrons or muons are misidentified as $\tauh$ candidates, predominantly $\PZ\to \Pe\Pe$ events in the $\etau$ channel and $\PZ\to \Pgm\Pgm$ events in the $\mutau$ channel,
rate uncertainties of 12 and 25\%, respectively, are allocated, based on the tag-and-probe method.

\subsection{Shape uncertainties}
The energy scales of the $\tauh$ candidate and the leading jet affect the shape of the $\st$
distribution, as well as the normalization of the signal and background processes.
The uncertainty is estimated by varying the $\tauh$ and jet energies within their respective
uncertainties and recomputing $\st$ after the final selection.
The uncertainty in the $\tauh$ energy scale amounts to 3\%\,\cite{Khachatryan:2015dfa}, whereas the variations due to the jet energy scale are in the 1--2\% range, depending on the jet $\pt$ and $\eta$\,\cite{CMS-JME-10-011}.

The uncertainty in the extrapolation of the $\tauh$ identification efficiency to higher $\pt$s is treated as a shape uncertainty. It is proportional to $\pt (\tauh)$ and has a value of  $+5\%$/$-35\%$ at $\pt (\tauh) = 1$\TeV.
The effects of the uncertainties due to the electron and muon energy scales are found to be negligible.
The probability of a jet being misidentified as a $\tauh$ candidate has been checked using a $\ttbar$ control region in data. The difference between data and simulated events is fit using a linear function, and its functional form, $1.2 - 0.004 [\max(120, \pt)\GeV]$, is considered as a one-sided shape uncertainty for the $\ltau$ channels.

In the $\tautau$ channel, additional shape uncertainties related to the fake-factor method are considered.
There are eight variations coming from factors such as the finite number of events in the samples, possible neglected effects in the fake factor determination, additional uncertainties in the correction of the SS to OS extrapolation, and the uncertainties in the background composition in the AR, which is estimated with simulated events. When added in quadrature, these additional uncertainties are of order 10\%.

Finally, uncertainties related to the finite number of simulated events, and to the limited number of events in data CRs, are taken into account. They are considered for all bins of the distributions that are used to extract the results. The binning of the histograms are adjusted such that the uncertainty, for a given bin, does not exceed 15\%.
They are uncorrelated
across the different samples, and across the bins of a single distribution.

\begin{table}[htbp]
\centering
\topcaption{Summary of systematic uncertainties in the signal acceptance and background estimate.
The uncertainties have been grouped into those affecting the normalization of distributions and those affecting the shape, and uncertainties marked with a * are treated as correlated among channels.}

\begin{tabular}{llP{7em}P{7em}P{7em}}
\hline
 & Systematic source & \multicolumn{3}{c}{Uncertainty}\\
 &  & $\etau$ & $\mutau$ & $\tautau$  \\
\hline
\multicolumn{5}{l}{{Normalization}} \\
& Luminosity * & 2.5\% & 2.5\% & 2.5\% \\
& Electron identification & 8\% & \NA & \NA \\
& Electron trigger & 2\% & \NA & \NA \\
& Muon identification & \NA & 2\%   & \NA \\
& Muon trigger & \NA & 2\%   & \NA \\
& $\tauh$ identification * & 5\% & 5\% & 10\% \\
& $\tauh$ trigger * & \NA & \NA & 10\% \\
& $\PQb$ tagging efficiency * & 3\% & 3\% & 3\% \\
& $\PQb$ tagging misidentification rate * & 5\% & 5\% & 5\% \\
& QCD multijet normalization & 30\% & 30\% & \NA \\
& $\wjet$ normalization & 30\% & 30\% & \NA \\
& $\PZ/\gamma*\to \ell \ell$ cross section * & 30\% & 30\% & 30\% \\
& \ttbar cross section *& 5.5\% & 5.5\%   & 5.5\% \\
& Diboson cross section *& 6\% & 6\%   & 6\% \\
& Single top quark cross section *& 5.5\% & 5.5\%  & 5.5\% \\
& $\Pe \to \tauh$ misidentification rate & 12\% & \NA  & \NA \\
& $\mu \to \tauh$ misidentification rate & \NA & 25\%  & \NA \\
\multicolumn{5}{l}{{Shape}} \\
& $\tauh$ energy scale * & \multicolumn{3}{c}{$\pm$3\%} \\
& $\tauh$ identification extrapolation * & \multicolumn{3}{c}{$+5\% \pt(\tauh)$ and $-35\% \pt(\tauh)$}   \\
& Jet energy scale *& \multicolumn{3}{c}{$\pm 1$ standard deviation\,\cite{CMS-JME-10-011}} \\
& Jet $\to \tauh$ misidentification rate *& \multicolumn{3}{c}{Described in the text (only $\ltau$ channels)} \\
& Fake-factor method & \multicolumn{3}{c}{Described in the text (only $\tautau$ channel)} \\
& Simulated sample size & \multicolumn{3}{c}{Statistical uncertainty in individual bins} \\ \hline
\end{tabular}
\label{table:uncertainties}
\end{table}

\section{Results}
\label{sec:result}

\begin{figure}[h!t]
\centering
\includegraphics[width=0.44\textwidth]{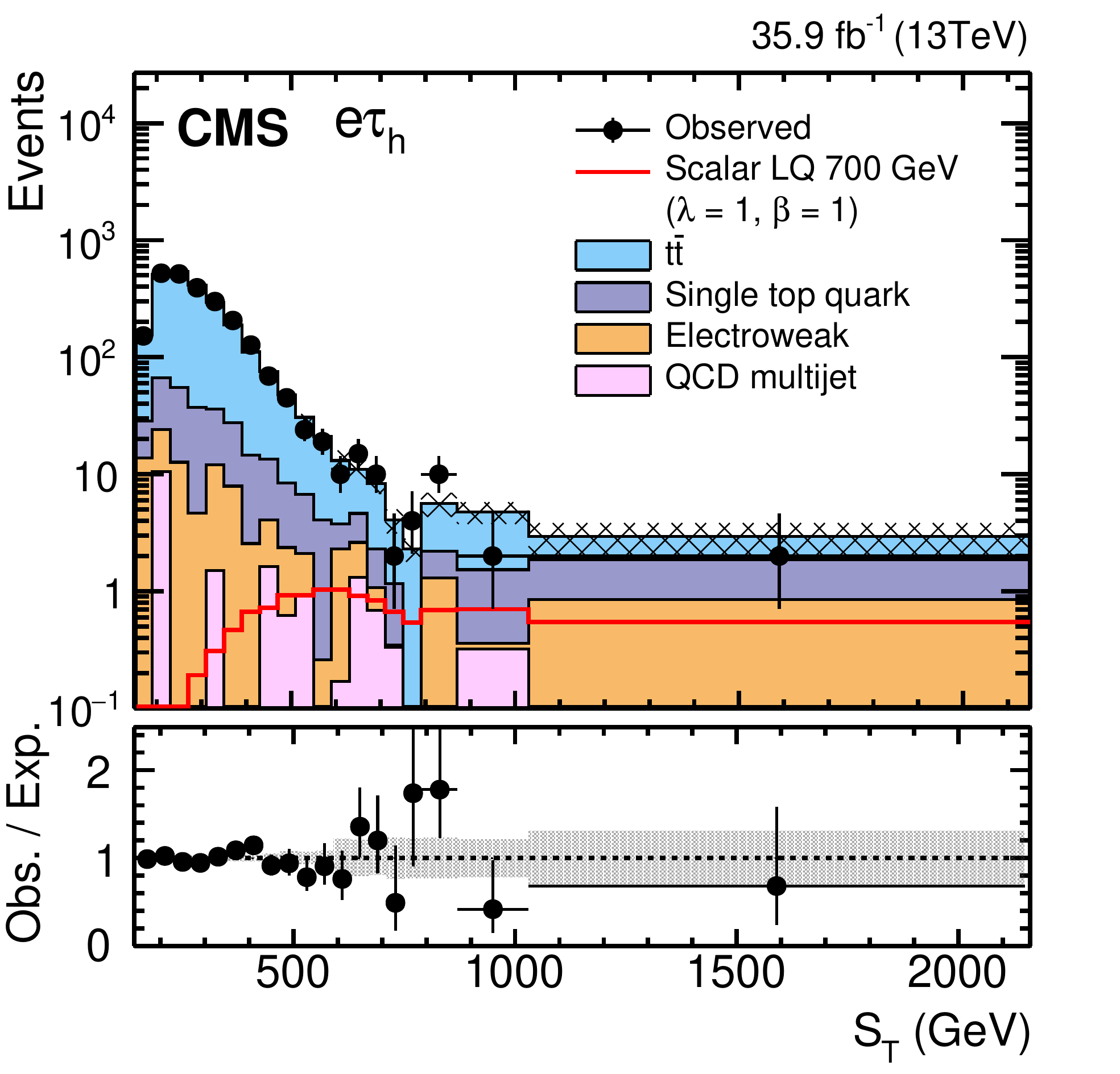}
\includegraphics[width=0.44\textwidth]{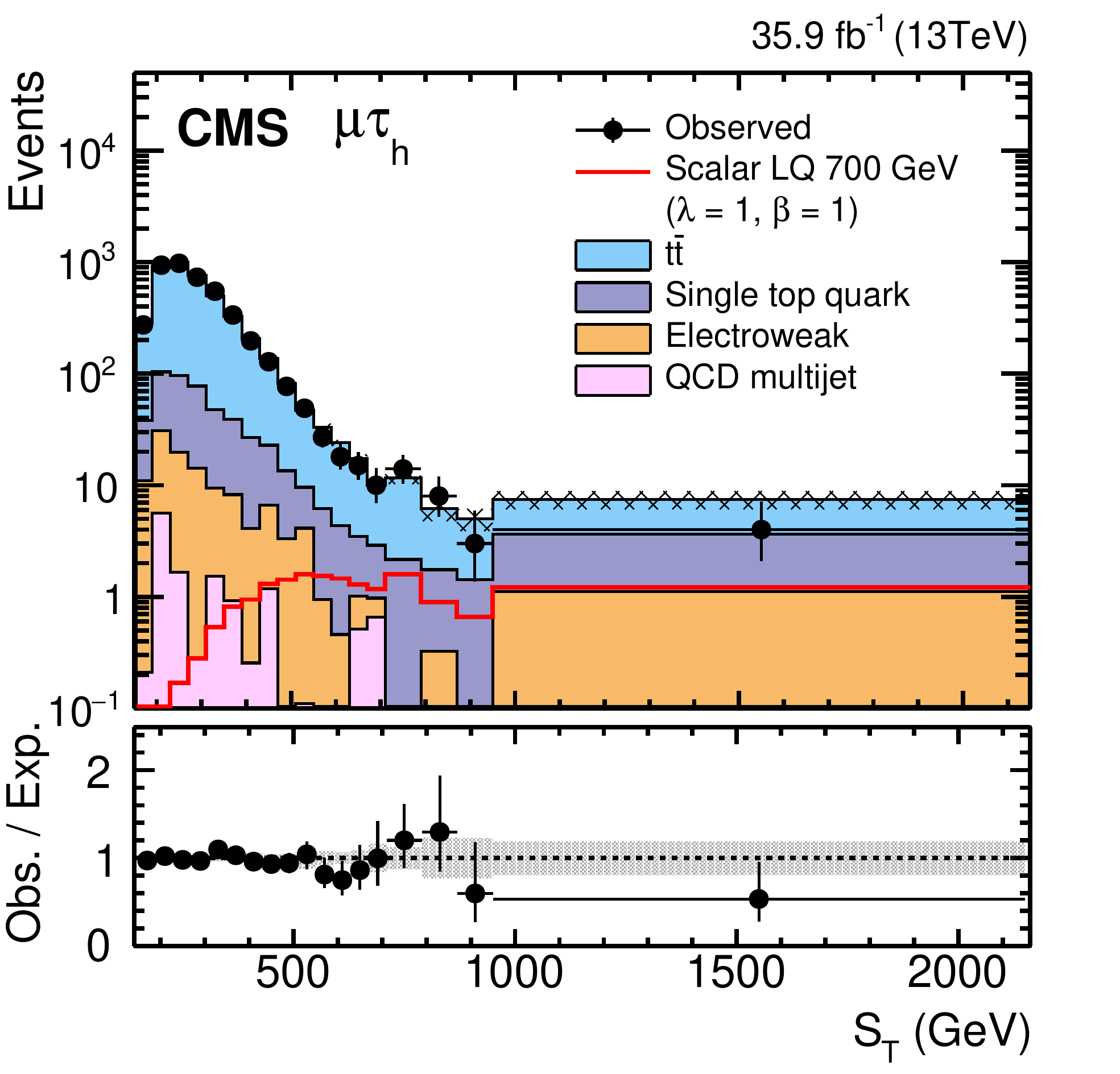}
\includegraphics[width=0.44\textwidth]{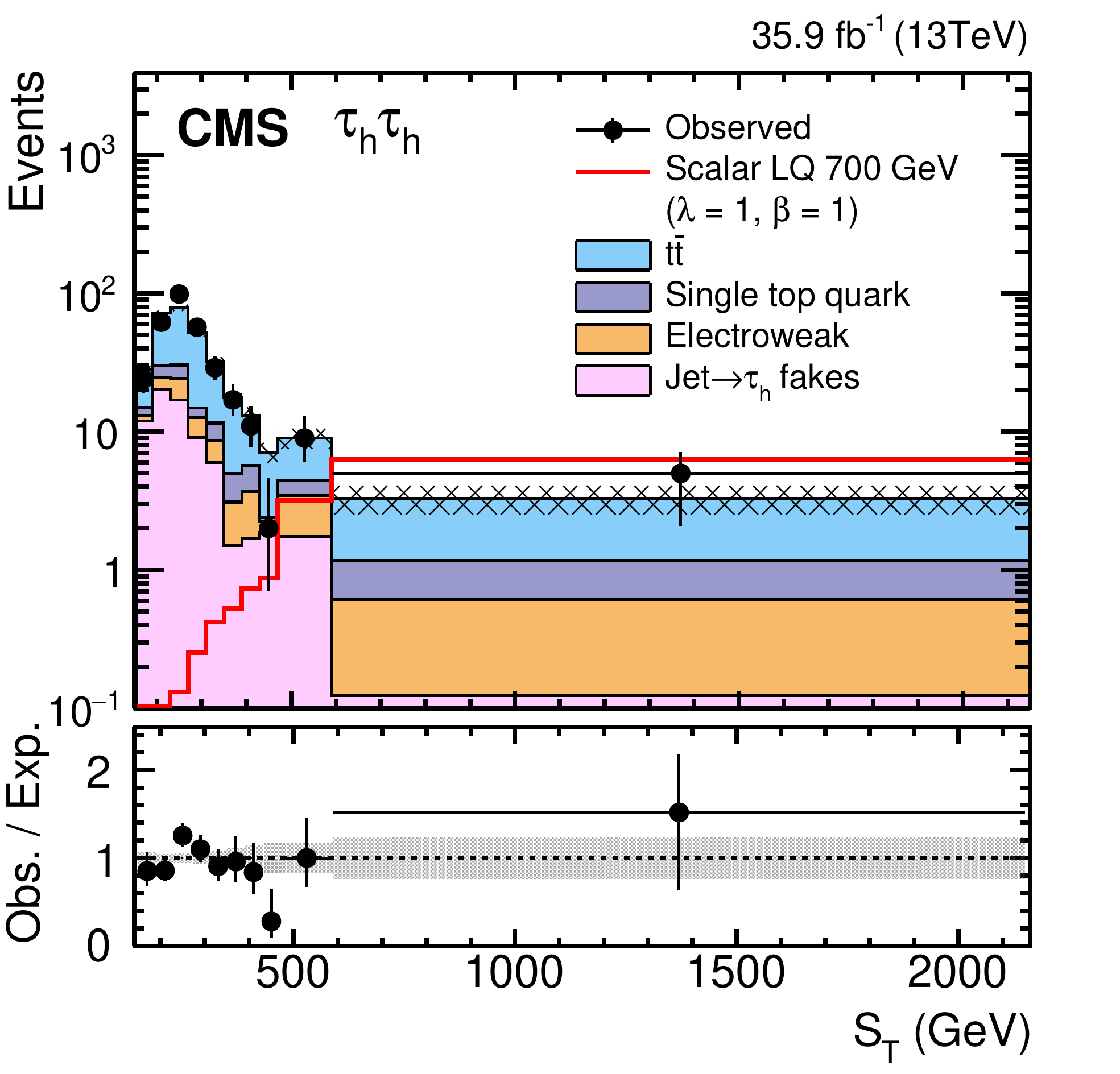}
\includegraphics[width=0.44\textwidth]{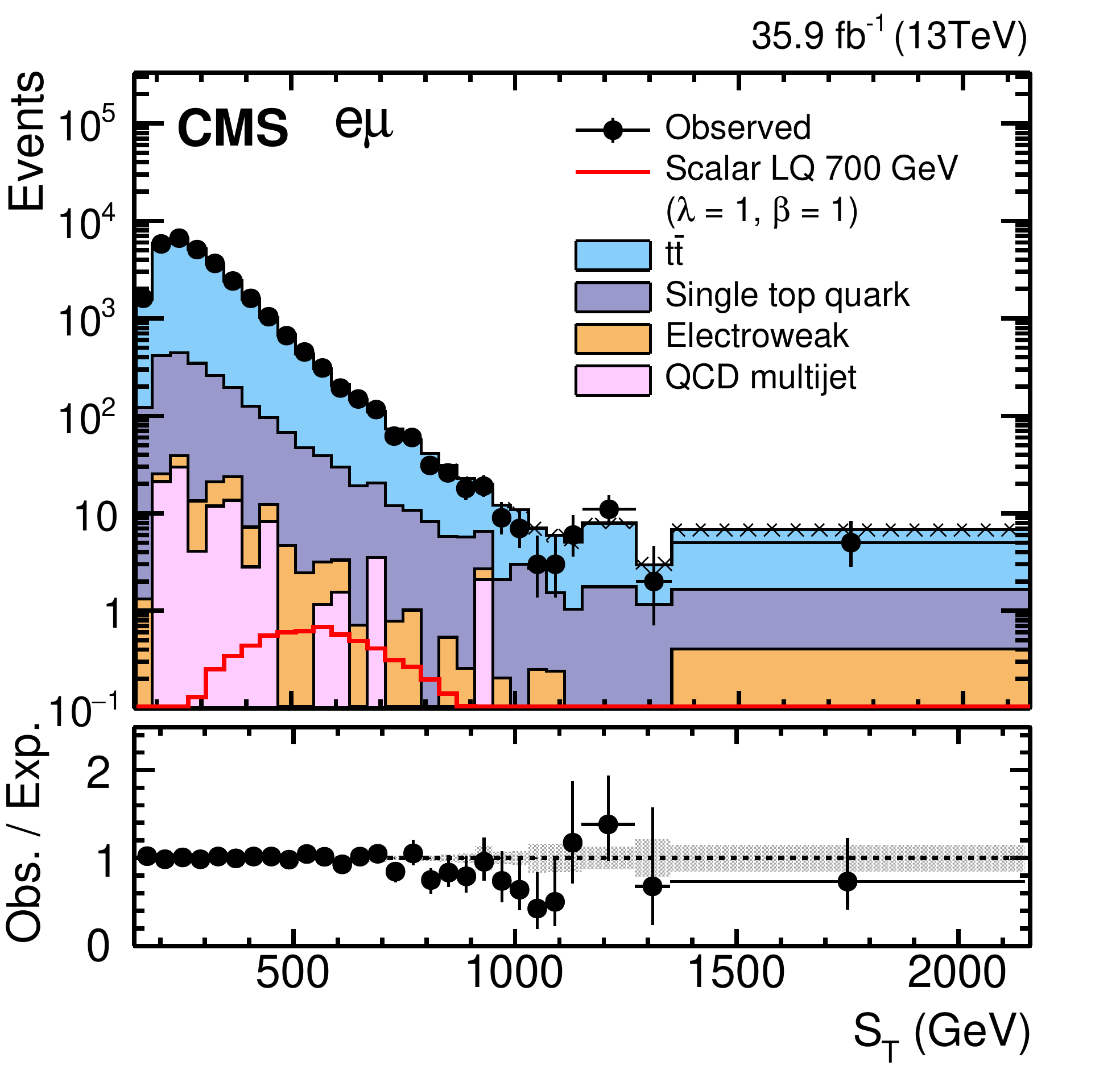}
\caption{Observed $\st$ distribution in the $\etau$ (upper left), $\mutau$ (upper right), and $\tautau$ (lower left) signal regions, as well as in the $\emu$ (lower right) control region, compared to the
expected SM background contributions. The distribution labeled ``electroweak'' contains the contributions from $\wjet$, $\zjet$, and diboson processes. The signal distributions for single-LQ production with mass 700\GeV are overlaid to illustrate the sensitivity. For the signal normalization, $\lambda=1$ and $\beta=1$ are assumed. The background uncertainty bands represent the sum in quadrature of statistical and systematic uncertainties obtained from the fit. The lower panels show the ratio between the observed and expected events in each bin. In all plots, the horizontal and vertical error bars on the data points represent the bin widths and the Poisson uncertainties, respectively.}
\label{fig:discriminant_final}
\end{figure}

Figure\,\ref{fig:discriminant_final} shows the $\st$ distributions after the combined fit to the $\etau$, $\mutau$, and $\tautau$ signal regions, as well as to the $\emu$ control region.
The background uncertainty bands on the histograms of simulated events represent the sum in quadrature of statistical and systematic uncertainties,
taking the full covariance matrix of all nuisance parameters into account. The dominant uncertainty in the background estimate comes from the limited event counts in simulated samples.
However, this uncertainty is unimportant for $\mLQ > 500\GeV$, where the mass limit is set, and the sensitivity is ultimately constrained by the size of the data sample.

Table\,\ref{tab:yield2} shows the event yields for a signal-enriched region with $\st > 500\GeV$, together with
event yields expected for a representative LQ signal with $\mLQ = 700\GeV$.

\begin{table}[htbp]

\centering
\topcaption{
Numbers of events observed in the $\etau$, $\mutau$, $\tautau$, and $\emu$ channels for $\st > 500\GeV$, compared to the background expectations and to the event yield expected for single-LQ processes with $\mLQ = 700\GeV$ ($\lambda=1$ and $\beta=1$). The ``electroweak'' background contains the contributions from $\wjet$, $\zjet$, and diboson processes. The uncertainties represent the sum in quadrature of statistical and systematic contributions, and are obtained using the binned maximum likelihood fit of the $\st$ distribution.}
\begin{tabular}{
  l
  r
  @{${}\pm{}$}
  r
  r
  @{${}\pm{}$}
  r
  r
  @{${}\pm{}$}
  r
  r
  @{${}\pm{}$}
  r
}

\hline
Process & \multicolumn{2}{c}{$\etau$} & \multicolumn{2}{c}{$\mutau$} & \multicolumn{2}{c}{$\tautau$} & \multicolumn{2}{c}{$\emu$} \\
\hline
\rule{0pt}{2.2ex}$\ttbar$ & 114.8 & 2.9 & 194.6 & 4.4 & 6.7 & 1.0 & 1895.2 & 14.4  \\
Single top quark & 23.2 & 2.2 & 36.6 & 2.6 & 1.5 & 0.5 & 263.4 & 6.8  \\
Electroweak & 9.1 & 2.3 &  10.9 & 3.1 & 2.2 & 1.0 & 16.0 & 2.4 \\
QCD multijet & 4.5 & 4.6 & 1.5 & 5.3 & 1.9 & 0.6 & 8.3 & 5.6  \\[\cmsTabSkip]
Total expected background & 151.6 & 6.3 & 243.6 & 8.0  &  12.3 & 1.7 & 2182.9 & 17.0  \\[\cmsTabSkip]
LQ signal ($\mLQ=700\GeV$, $\lambda = 1$, $\beta = 1$) & 8.8 & 0.3 & 12.9 & 0.4 & 9.5 & 1.2 & 4.9 & 0.2 \\[\cmsTabSkip]
Observed data &   \multicolumn{2}{c}{\hspace{0.35cm} 143} & \multicolumn{2}{c}{\hspace{0.35cm} 225} & \multicolumn{2}{c}{\hspace{0.15cm} 14} & \multicolumn{2}{c}{\hspace{0.35cm} 2147} \\ \hline
\end{tabular}
\label{tab:yield2}
\end{table}

The data are consistent with the background-only (SM) hypothesis.
In the $\tautau$ channel, one bin at around 250\GeV shows a slight excess in data, corresponding to two standard deviations. This, however, has little impact on the results, as the sensitivity is dominated by the $\st$ tail, rather than the main part of the distribution.

\begin{figure}[h!t]
\centering
\includegraphics[width=0.6\textwidth]{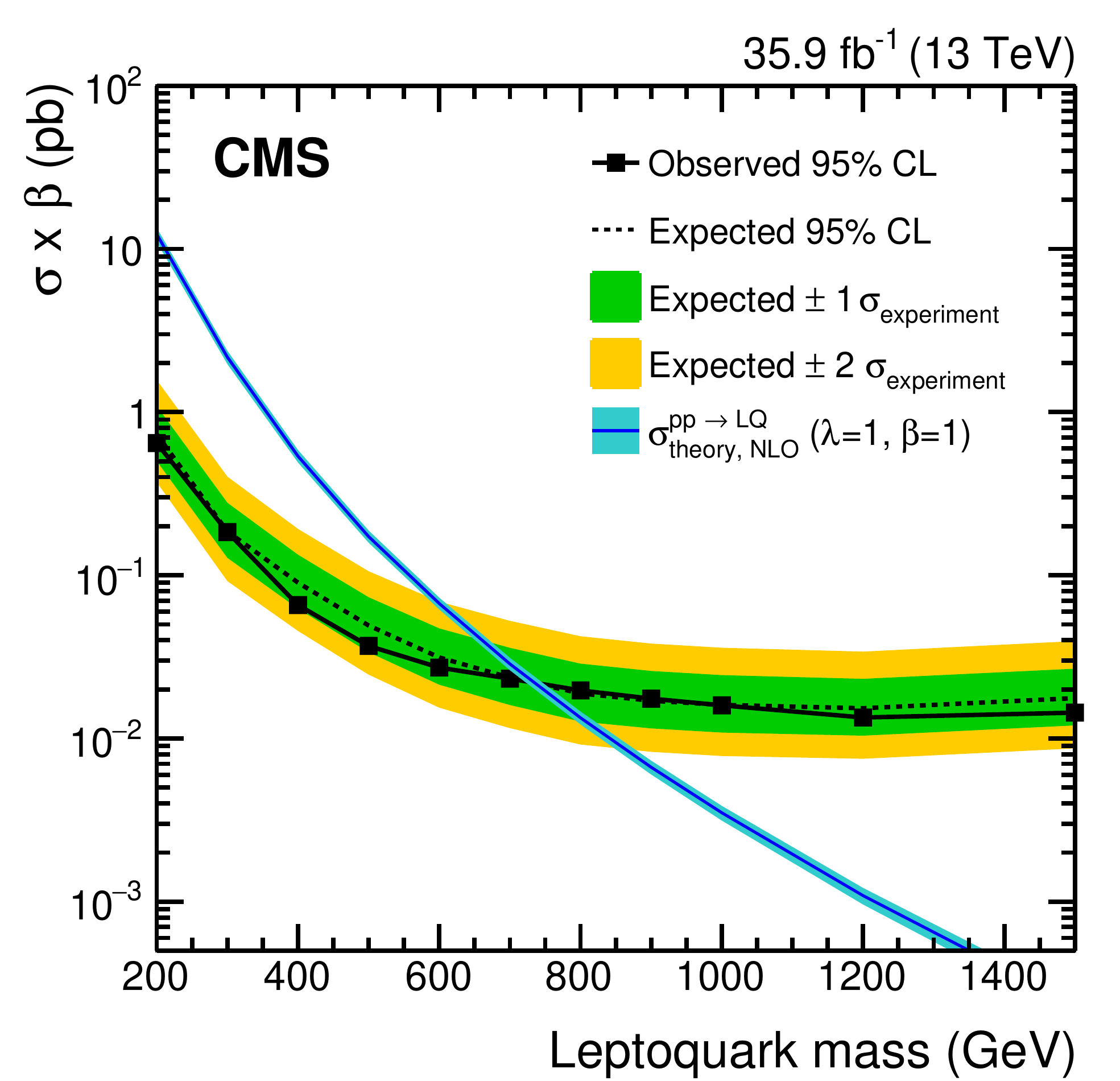}
\caption{Observed (black solid) and expected (black dotted) limits at 95\% confidence level on the product of cross section $\sigma$ and branching fraction $\beta$, obtained from the combination of the $\etau$, $\mutau$, and $\tautau$ signal regions, as well as from the $\emu$ control region, as a function of the LQ mass. The green and yellow bands represent the one and two standard deviation uncertainties in the expected limits. The theory prediction is indicated by the blue solid line,
together with systematic uncertainties due to the choice of PDF and renormalization and factorization scales\,\cite{Dorsner:2018ynv}, indicated by the blue band.}
\label{fig:limit_comb}
\end{figure}

We set an upper limit on the cross section times branching fraction $\beta$ as a function of $\mLQ$, by using the asymptotic \CLs modified frequentist criterion\,\cite{Cowan:2010js, Junk:1999kv, 0954-3899-28-10-313, CMS-NOTE-2011-005}.
Figure\,\ref{fig:limit_comb} shows the observed and expected upper limits at 95\% confidence level.
The blue solid line corresponds to the theoretical cross sections\,\cite{Dorsner:2018ynv}, calculated with $\lambda = 1$ and $\beta = 1$. The intersection of the blue and the black lines determines the lower limit on $\mLQ$.
Assuming $\lambda = 1$ and $\beta = 1$, third-generation scalar LQs with masses below 740\GeV are excluded at 95\% confidence level, to be compared with an expected lower limit of 750\GeV.

The sensitivity of the analysis is dominated by the $\tautau$ channel,
followed by the $\mutau$ channel, and then the $\etau$ channel. The better sensitivity in the $\tautau$ channel comes from the larger branching fraction of $\mathcal{B}(\tau\tau \to \tautau) = 42$\%, compared to $\mathcal{B}(\tau\tau \to \mutau) = \mathcal{B}(\tau\tau \to \etau) = 21$\%.
Furthermore, the $\ltau$ channels are contaminated by $\ttbar\to \PW\PW \PQb\PQb \to \ltau \nu\nu \PQb\PQb$ process, in addition to the $\ttbar\to \PW\PW \PQb\PQb \to\tau_{\ell} \tauh \nu\nu \PQb\PQb$ background, which is not the case for the $\tautau$ channel. Here, $\tau_\ell$ denotes a leptonically-decaying $\tau$ lepton.

\begin{figure}[h!t]
\centering
\includegraphics[width=0.6\textwidth]{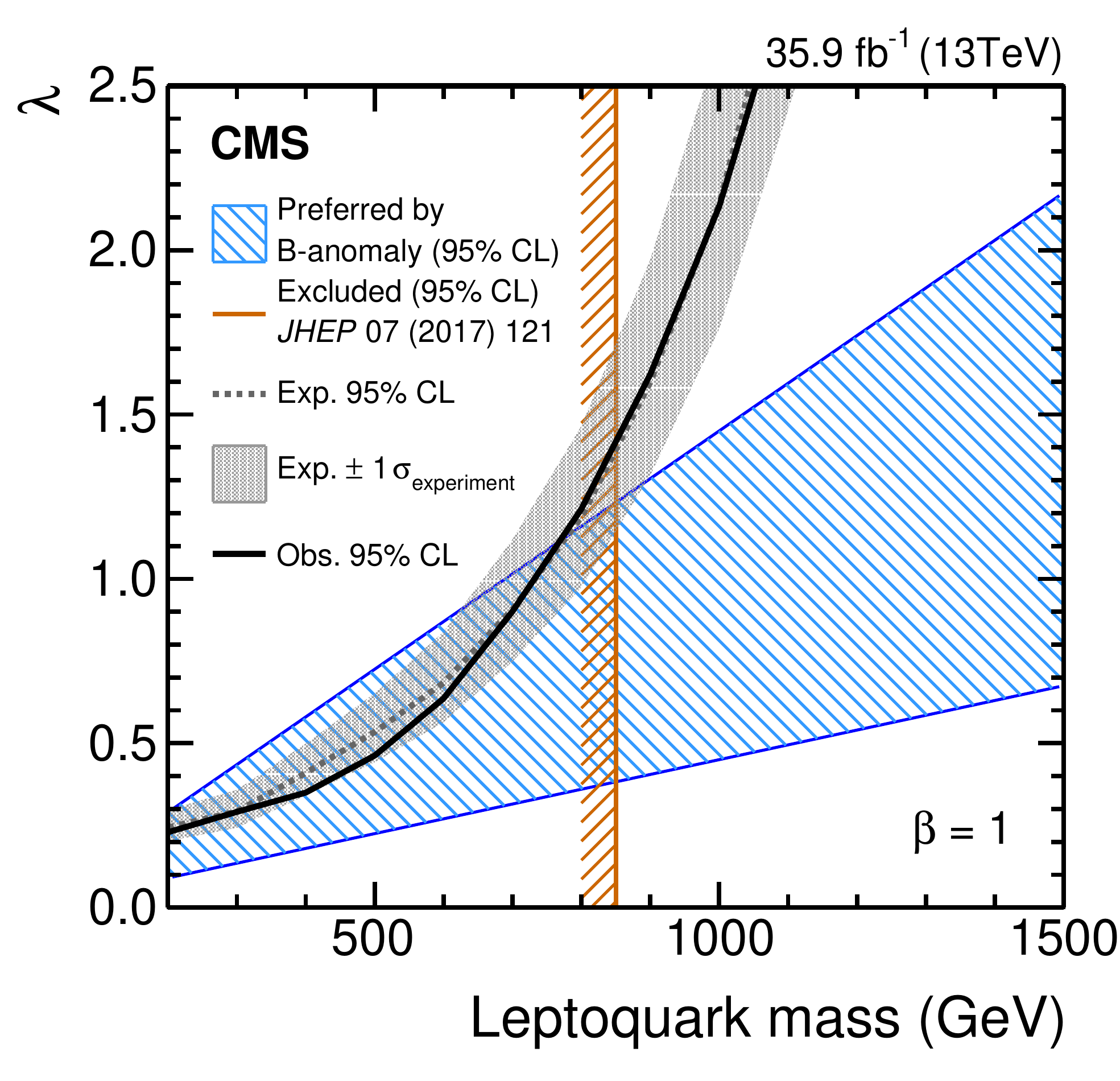}
\caption{Expected and observed exclusion limits at 95\% confidence level on the Yukawa coupling $\lambda$ at the LQ-lepton-quark vertex, as a function of the LQ mass. A unit branching fraction $\beta$ of the LQ to a $\tau$ lepton and a bottom quark is assumed. The orange vertical line indicates the limit obtained from a search for pair-produced LQs decaying to $\ltau\PQb\PQb$\,\cite{Sirunyan:2017yrk}. The left-hand side of the dotted (solid) line shows the expected (observed) exclusion region for the present analysis. The gray band shows $\pm 1$ standard deviations of the expected exclusion limit. The region with diagonal blue shading shows the parameter space preferred by one of the models proposed to explain anomalies observed in $\PB$ physics\,\cite{Buttazzo:2017ixm}.}
\label{fig:lambda_mass}
\end{figure}

Since the single-LQ signal cross section scales with $\lambda^2$, it is straightforward to recast the results presented in Fig.\,\ref{fig:limit_comb} in terms of expected and observed upper limits on $\lambda$ as a function of $\mLQ$, as shown in Fig.\,\ref{fig:lambda_mass}. Values of $\lambda$ up to 2.5 are considered, such that the width of the LQ signal stays narrow and to satisfy constraints from electroweak precision measurements\,\cite{Dorsner:2016wpm}. Here we have made the assumption that the shape of the $\st$ distribution does not depend on $\lambda$ over the range of $\lambda$ used in the analysis.  This assumption has been verified using simulated events.
The blue band shows the preferred parameter space (95\% CL) for the scalar LQ preferred by the $\PB$ physics anomalies: $\lambda = (0.95 \pm 0.50) \mLQ (\TeVns{})$\,\cite{Buttazzo:2017ixm}.
The plot also shows the limit from the pair-produced LQ search overlaid as an orange vertical line, which does not depend on $\lambda$, as discussed in Section\,\ref{sec:intro}.
For values of $\lambda > 1.4$, the mass limit obtained by this analysis exceeds that of the search considering pair production\,\cite{Sirunyan:2017yrk} and provides the best upper limit on $\mLQ$ of the third-generation scalar LQ decaying to a $\tau$ lepton and a bottom quark. For $\lambda = 2.5$ and $\beta = 1$, the observed and expected lower limits on mass are both 1050\GeV. This result, together with the pair-produced search, begins to constrain the region of parameter space implied by the $\PB$ physics anomalies.

\section{Summary}
A search for singly produced third-generation scalar leptoquarks, each decaying to a $\tau$ lepton and a bottom quark has been presented.
The final state of an electron or a muon plus one hadronically decaying $\tau$ lepton and the final state with two hadronically decaying $\tau$ leptons are explored. In all final states at least one energetic jet identified as originating from a bottom quark is required. The search is based on a data sample of proton-proton collisions at a center-of-mass energy of 13\TeV recorded by the CMS detector, corresponding to an integrated luminosity of 35.9\fbinv. The data are found to be in agreement with the standard model predictions.
Upper limits as a function of the leptoquark mass are set on the third-generation scalar leptoquark production cross section.
Results are compared with theoretical predictions to obtain lower limits on the leptoquark mass.
Assuming the leptoquark always decays to a $\tau$ lepton and a bottom quark with unit Yukawa coupling $\lambda = 1$, third-generation scalar leptoquarks with mass below 740\GeV are excluded at 95\% confidence level.
Mass limits are also placed as a function of $\lambda$. For values of $\lambda > 1.4$, the mass limit obtained by this analysis exceeds that of the search considering pair production and provides the best upper limit. For $\lambda = 2.5$, leptoquarks are excluded in the mass range up to 1050\GeV.
This is the first time that limits have been presented in the $\lambda$ versus mass plane, allowing the results to be considered in the preferred parameter space of models that invoke third-generation leptoquarks to explain anomalies observed in $\PB$ hadron decays.
These results thus demonstrate the important potential of single
leptoquark production studies to complement pair production constraints
on such models, as additional data become available.

\begin{acknowledgments}
We congratulate our colleagues in the CERN accelerator departments for the excellent performance of the LHC and thank the technical and administrative staffs at CERN and at other CMS institutes for their contributions to the success of the CMS effort. In addition, we gratefully acknowledge the computing centers and personnel of the Worldwide LHC Computing Grid for delivering so effectively the computing infrastructure essential to our analyses. Finally, we acknowledge the enduring support for the construction and operation of the LHC and the CMS detector provided by the following funding agencies: BMWFW and FWF (Austria); FNRS and FWO (Belgium); CNPq, CAPES, FAPERJ, FAPERGS, and FAPESP (Brazil); MES (Bulgaria); CERN; CAS, MoST, and NSFC (China); COLCIENCIAS (Colombia); MSES and CSF (Croatia); RPF (Cyprus); SENESCYT (Ecuador); MoER, ERC IUT, and ERDF (Estonia); Academy of Finland, MEC, and HIP (Finland); CEA and CNRS/IN2P3 (France); BMBF, DFG, and HGF (Germany); GSRT (Greece); NKFIA (Hungary); DAE and DST (India); IPM (Iran); SFI (Ireland); INFN (Italy); MSIP and NRF (Republic of Korea); LAS (Lithuania); MOE and UM (Malaysia); BUAP, CINVESTAV, CONACYT, LNS, SEP, and UASLP-FAI (Mexico); MBIE (New Zealand); PAEC (Pakistan); MSHE and NSC (Poland); FCT (Portugal); JINR (Dubna); MON, RosAtom, RAS, RFBR, and NRC KI (Russia); MESTD (Serbia); SEIDI, CPAN, PCTI, and FEDER (Spain); Swiss Funding Agencies (Switzerland); MST (Taipei); ThEPCenter, IPST, STAR, and NSTDA (Thailand); TUBITAK and TAEK (Turkey); NASU and SFFR (Ukraine); STFC (United Kingdom); DOE and NSF (USA). and SFFR (Ukraine); STFC (United Kingdom); DOE and NSF (USA).

\hyphenation{Rachada-pisek} Individuals have received support from the Marie-Curie program and the European Research Council and Horizon 2020 Grant, contract No. 675440 (European Union); the Leventis Foundation; the A. P. Sloan Foundation; the Alexander von Humboldt Foundation; the Belgian Federal Science Policy Office; the Fonds pour la Formation \`a la Recherche dans l'Industrie et dans l'Agriculture (FRIA-Belgium); the Agentschap voor Innovatie door Wetenschap en Technologie (IWT-Belgium); the F.R.S.-FNRS and FWO (Belgium) under the ``Excellence of Science - EOS" - be.h project n. 30820817; the Ministry of Education, Youth and Sports (MEYS) of the Czech Republic; the Lend\"ulet (``Momentum") Program and the J\'anos Bolyai Research Scholarship of the Hungarian Academy of Sciences, the New National Excellence Program \'UNKP, the NKFIA research grants 123842, 123959, 124845, 124850 and 125105 (Hungary); the Council of Science and Industrial Research, India; the HOMING PLUS program of the Foundation for Polish Science, cofinanced from European Union, Regional Development Fund, the Mobility Plus program of the Ministry of Science and Higher Education, the National Science Center (Poland), contracts Harmonia 2014/14/M/ST2/00428, Opus 2014/13/B/ST2/02543, 2014/15/B/ST2/03998, and 2015/19/B/ST2/02861, Sonata-bis 2012/07/E/ST2/01406; the National Priorities Research Program by Qatar National Research Fund; the Programa Estatal de Fomento de la Investigaci{\'o}n Cient{\'i}fica y T{\'e}cnica de Excelencia Mar\'{\i}a de Maeztu, grant MDM-2015-0509 and the Programa Severo Ochoa del Principado de Asturias; the Thalis and Aristeia programs cofinanced by EU-ESF and the Greek NSRF; the Rachadapisek Sompot Fund for Postdoctoral Fellowship, Chulalongkorn University and the Chulalongkorn Academic into Its 2nd Century Project Advancement Project (Thailand); the Welch Foundation, contract C-1845; and the Weston Havens Foundation (USA).
\end{acknowledgments}

\bibliography{auto_generated}
\cleardoublepage \appendix\section{The CMS Collaboration \label{app:collab}}\begin{sloppypar}\hyphenpenalty=5000\widowpenalty=500\clubpenalty=5000\vskip\cmsinstskip
\textbf{Yerevan Physics Institute, Yerevan, Armenia}\\*[0pt]
A.M.~Sirunyan, A.~Tumasyan
\vskip\cmsinstskip
\textbf{Institut f\"{u}r Hochenergiephysik, Wien, Austria}\\*[0pt]
W.~Adam, F.~Ambrogi, E.~Asilar, T.~Bergauer, J.~Brandstetter, M.~Dragicevic, J.~Er\"{o}, A.~Escalante~Del~Valle, M.~Flechl, R.~Fr\"{u}hwirth\cmsAuthorMark{1}, V.M.~Ghete, J.~Hrubec, M.~Jeitler\cmsAuthorMark{1}, N.~Krammer, I.~Kr\"{a}tschmer, D.~Liko, T.~Madlener, I.~Mikulec, N.~Rad, H.~Rohringer, J.~Schieck\cmsAuthorMark{1}, R.~Sch\"{o}fbeck, M.~Spanring, D.~Spitzbart, A.~Taurok, W.~Waltenberger, J.~Wittmann, C.-E.~Wulz\cmsAuthorMark{1}, M.~Zarucki
% ...
\vskip\cmsinstskip
% ...
1:  Also at Vienna University of Technology, Vienna, Austria\\
% ...
\end{sloppypar}
\end{document}